\documentclass[aps,pra,reprint,showpacs,preprintnumbers,amsmath,amssymb]{revtex4-1} 
\usepackage{amsmath}
\usepackage{graphicx}
\usepackage{amssymb}
\usepackage{textcomp}
\usepackage{color}
\usepackage{xcolor}
\usepackage{float}
\usepackage{natbib}
\usepackage{dcolumn}
\usepackage{ragged2e} 
\usepackage{color,soul} 
\usepackage{array}
\usepackage{dcolumn}
\usepackage{url}
\usepackage{multirow}
\usepackage{appendix}

\usepackage[normalem]{ulem}

\usepackage{soul}

\usepackage{caption}
\usepackage{subcaption}

\usepackage[colorlinks=true,linkcolor=black,citecolor=black,urlcolor=blue,bookmarks=false,hypertexnames=false]{hyperref}
\newcolumntype{.}{D{.}{.}{-1}}
\newcolumntype{L}[1]{>{\raggedright\let\newline\\\arraybackslash\hspace{0pt}}m{#1}}
\newcolumntype{C}[1]{>{\centering\let\newline\\\arraybackslash\hspace{0pt}}m{#1}}
\newcolumntype{R}[1]{>{\raggedleft\let\newline\\\arraybackslash\hspace{0pt}}m{#1}}

\begin{document}

\preprint{}

\title{Two-photon optical shielding of collisions between ultracold polar molecules}

\author{Charbel Karam}
\affiliation{Universit$\acute{\text e}$ Paris-Saclay, CNRS, Laboratoire Aim$\acute{\text e}$ Cotton, Orsay, 91400, France}
\author{Romain Vexiau}
\affiliation{Universit$\acute{\text e}$ Paris-Saclay, CNRS, Laboratoire Aim$\acute{\text e}$ Cotton, Orsay, 91400, France}
\author{Nadia Bouloufa-Maafa}
\affiliation{Universit$\acute{\text e}$ Paris-Saclay, CNRS, Laboratoire Aim$\acute{\text e}$ Cotton, Orsay, 91400, France}
\author{Olivier Dulieu}
\affiliation{Universit$\acute{\text e}$ Paris-Saclay, CNRS, Laboratoire Aim$\acute{\text e}$ Cotton, Orsay, 91400, France}

\author{Maxence Lepers}
\affiliation{ Laboratoire Interdisciplinaire Carnot de Bourgogne, CNRS, Univ. Bourgogne Franche-Comt$\acute{\text e}$, Cedex F-21078 Dijon, France}

\author{Mara Meyer zum Alten Borgloh}
\affiliation{Institut für Quantenoptik, Leibniz Universität Hannover, 30167 Hannover, Germany}

\author{Silke Ospelkaus}
\affiliation{Institut für Quantenoptik, Leibniz Universität Hannover, 30167 Hannover, Germany}

\author{Leon Karpa}
\affiliation{Institut für Quantenoptik, Leibniz Universität Hannover, 30167 Hannover, Germany}

\date{\today}

\begin{abstract}

We propose a method to engineer repulsive long-range interactions between ultracold ground-state molecules using optical fields, thus preventing short-range collisional losses. It maps the microwave coupling recently used for collisional shielding onto a two-photon transition, and takes advantage of optical control techniques. In contrast to one-photon optical shielding [Phys. Rev. Lett. 125, 153202 (2020)], this scheme avoids heating of the molecular gas due to photon scattering. The proposed protocol, exemplified for $^{23}$Na$^{39}$K, should be applicable to a large class of polar diatomic molecules.

\end{abstract}
\maketitle

%%%%%%%%%%%%%%%%%%%%%%%%%%%%%%%%%%%%%%%%%%%%%%%%%%%%%%%%%

\section{Introduction}

The full understanding and modelling of few-body systems remains a long-standing challenge in several areas of science, like for instance in quantum physics. The possibility to create and manipulate dilute gases at ultracold temperatures, composed of particles with kinetic energies $E=k_BT \ll 1$~mK, opened novel opportunities in this respect. The growing availability of quantum gases of ultracold polar molecules (\textit{i.e.} possessing a permanent electric dipole moment (PEDM) in their own frame) in several labs revealed a very peculiar situation in the context of few-body physics: at ultracold energies, two such molecules in their absolute ground level (\textit{i.e.} in the lowest rovibrational and hyperfine level of their electronic ground state) collide  with a universal collisional rate, even if they have no inelastic or reactive energetically allowed channels, so that they leave the molecular trap with a short characteristic time. Such a four-body system, which may look rather simple at first glance, is not yet fully characterized. It is currently interpreted as a "sticky" four-body complex \cite{mayle2013,croft2014}, with a huge density of states, for which various statistical models have been developed \cite{christianen2019a,christianen2019b,jachymski2022}. However, the experimental observations reported up to now regarding the molecular loss rates \cite{takekoshi2014, guo2016,park2015, voges2020} cannot yet be consistently reproduced by these models \cite{gregory2020,liu2020,bause2021,gersema2021}.

Instead of attempting to fully describe this four-body system, with the aim of identifying the exact cause of the universal loss rate, one can design protocols where molecules would simply not reach short distances in the course of their collision. Several options have been proposed and experimentally demonstrated, based on the modification of the long-range interaction (LRI) between molecules using static electric fields \cite{quemener2010a,wang2015,quemener2016,matsuda2020,li2021} or microwave (mw) fields \cite{,schindewolf2022,lassabliere2018,karman2018,anderegg2021,bigagli2023}, in order to "shield" their collisions. In a previous paper \cite{xie2020} we proposed an alternative way to engineer LRIs using a laser with a frequency blue-detuned from the one of a suitable molecular rovibronic transition. Such a one-photon optical shielding (1-OS), inspired from previous works on cold atoms \cite{marcassa1994,zilio1996}, results in the laser-induced coupling of the attractive collisional entrance channel to a repulsive one, thus preventing the molecules from reaching short distances, and from creating a sticky complex. One limitation of the 1-OS could be the heating of the molecular quantum gas due to the continuous scattering of off-resonant photons of the 1-OS laser.

\begin{figure}
    \centering
    \includegraphics[scale=0.7]{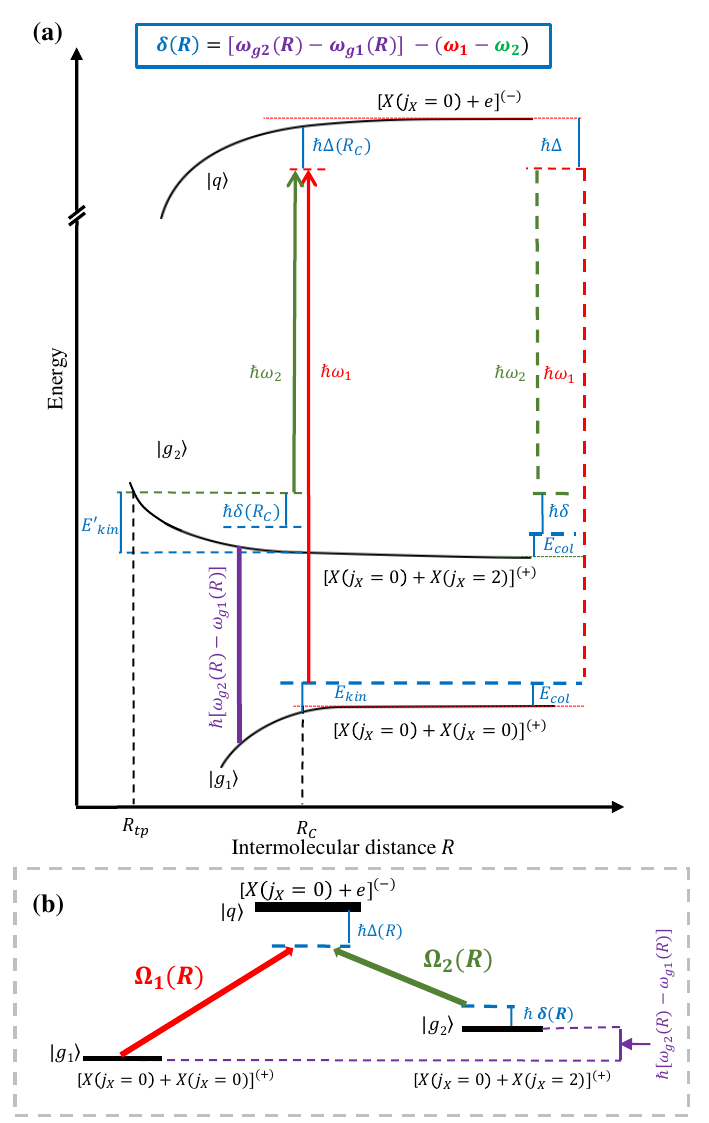}
    \caption{Schematic view of the proposed two-photon optical shielding (energies not to scale). (a) Sketch of the long-range adiabatic potential energy curves (PECs) in the space-fixed frame, describing the interaction between two polar molecules in the lowest vibrational level $v_X=0$ and rotational level $j_X=0$ or $j_X=2$ of their electronic ground state ($X$), and a ($v_X=0, j_X=0$) ground state molecule and a molecule in an electronically-excited one in state $e $. The sign at the asymptotes denotes the total parity (+) or (-) of the pair states. The photon energies $\hbar \omega_1$ and $\hbar \omega_2$ are represented by the vertical dashed lines, with their  detunings $\hbar \Delta$ and $\hbar \delta$. When two $X(j_X=0)$ molecules collide with an initial energy $E_{col}$ in the attractive pair state $|g_1\rangle$, they acquire a kinetic energy $E_{kin}$, and undergo a two-photon transition (solid vertical arrows) toward the $|g_2\rangle$ pair state at the Condon point $R = R_C$ (with kinetic energy $E'_{kin}$), via the excited pair state $|q\rangle$. Due to the $R$-variation of the potential energies $\hbar \omega_{g_1}$, $\hbar \omega_{g_2}$, and $\hbar \omega_{q}$, $\hbar \Delta$ and $\hbar \delta$ also vary with $R$. The $|g_2\rangle$ state has a repulsive PEC ($R_{tp}$ is the turning point corresponding to $E'_{kin}$).(b) The three-level coupling scheme at a given intermolecular distance $R$. }
    \label{fig1}
\end{figure}

\begin{figure}
    \centering
    \includegraphics[scale=0.9]{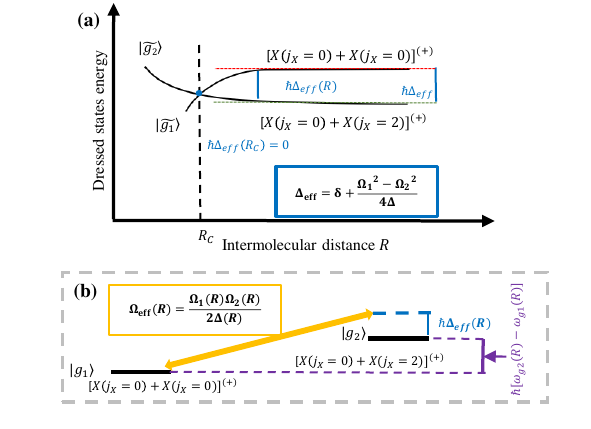}
    \caption{(a) The PECs of Fig.\ref{fig1} (a)  in the dressed-state framework involving the states $|\tilde{g}_1\rangle$ and $|\tilde{g}_2\rangle$: $R_C$ is the crossing point between the attractive and repulsive PECs, resulting from the effective detuning $\Delta_{\mathrm{eff}}$ (see text) at infinity.
    (b) The effective two-level scheme after applying the unitary transformation and the adiabatic elimination on the Hamiltonian (see text) at a given $R$.}
    \label{fig2}
\end{figure}

In this paper we propose a two-photon optical shielding (2-OS) scheme, aiming at overcoming the above limitation, while mapping the case of the microwave (mw) shielding (Fig.\ref{fig1}). As described below, such a scheme combines the best features of the 1-OS (no restriction for the field polarization, convenient laser power, tunability, geometrical versatility, broad compatibility) and mw shielding (no spontaneous emission or photon scattering). The scheme relies on coupling three molecular states $|g_1 \rangle$, $|q \rangle$ and $|g_2 \rangle$ of the collisional complex, with $R$-dependent energies (\textit{i.e.} the long-range potential energy curves (PECs) of Fig.\ref{fig1}a) via a two-photon transition from $|g_1 \rangle$ to $|g_2 \rangle$ occurring preferentially at the intermolecular distance $R=R_C$ (Fig.~\ref{fig1}b). In the dressed state picture (Fig.~\ref{fig2}a), this maps the mw shielding scheme \cite{karman2018,lassabliere2018} onto an effective optical coupling of the states $|\tilde{g}_1 \rangle$ and $|\tilde{g}_2 \rangle$ (\textit{i.e.} the dressed states).

\section{Model}
\subsection{Interaction Hamiltonian}\label{Interaction Hamiltonian}
We consider the Hamiltonian of the molecular pair $\hat{H}^I$ at a given $R$, namely, including the PECs in the space-fixed (SF) frame and ignoring the kinetic energy in this static point of view, its matrix $H^I(R)$ being expressed in the ($|g_1\rangle$, $|g_2\rangle$, $|q\rangle$) basis (see Fig. 1)
\begin{equation} 
H^I(R) = \\ 
\hbar
\begin{pmatrix}
\omega_{g_1}(R) & 0 & \Omega_1(R) \cos{\omega_1 t} \\
0 & \omega_{g_2}(R) & \Omega_2(R) \cos{\omega_2 t} \\
\Omega_1(R) \cos{\omega_1 t} & \Omega_2(R) \cos{\omega_2 t} & \omega_q(R)
\end{pmatrix},
\label{eq:H_3level}
\end{equation}
where $\hbar \omega_{g_1}(R)$, $\hbar \omega_{g_2}(R)$, $\hbar \omega_{q}(R)$ are the PECs of $|g_1\rangle$, $|g_2\rangle$, $|q\rangle$, respectively, and $\hbar \omega_1$ and $\Omega_1(R)$ (resp. $\hbar \omega_2$ and $\Omega_2(R)$) the energy and Rabi frequency of laser 1 (resp. laser 2) defined as:
\begin{equation}
\Omega_i(R)/(2 \pi) \mathrm{(MHz)}=  35.12 \times \sqrt{I_i \mathrm{(W/cm^2)}} \times d_{g_i q}(R) \mathrm{(ea_0)}    
\end{equation}
 with $i=1$ or $2$, $I_i$ the intensity of laser $i$, and $d_{g_i q}(R)$ the transition dipole moment between state $g_i$ and $q$.  
Following a standard approach for the treatment of a three-level system \cite{fleischhauer2005}, we apply a unitary transformation $\displaystyle{i \hbar \frac{d \hat{U}}{dt}\hat{U}^{-1} + \hat{U}\hat{H}^I\hat{U}^{-1}}$ that makes $\hat{H}^I$ time-independent by working in the so-called rotating frame, with
\begin{equation}
  \hat{U}= 
\begin{pmatrix}
e^{-i \omega_1 t} & 0 & 0 \\
0 & e^{-i \omega_2 t} & 0 \\
0 & 0 & 1
\end{pmatrix}.
\end{equation}
After neglecting the rapidly varying terms using the rotating wave approximation, we obtain the new Hamiltonian matrix $\tilde{H}^I(R)$ written in the basis of the dressed states ($|\tilde{g}_1\rangle, |\tilde{g}_2\rangle, |\tilde{q}\rangle$)
\begin{equation} 
\tilde{H}^I(R) = \hbar
\begin{pmatrix}
0 & 0 & \Omega_1(R)/2 \\
0 & \delta(R) & \Omega_2(R)/2 \\
\Omega_1(R)/2 & \Omega_2(R)/2 & \Delta(R)
\end{pmatrix}_{ |\tilde{g}_1\rangle, |\tilde{g}_2\rangle, |\tilde{q}\rangle},
\label{eq:Hdressed}
\end{equation}
where $\Delta(R)= [\omega_{q}(R)-\omega_{g_1}(R)] - \omega_{1}$ is the detuning of the laser coupling the initial state $|{g_1}\rangle$ to the intermediate electronically excited state $|{q}\rangle$, and $\delta(R)=[ \omega_{g_2}(R)-\omega_{g_1}(R)]-(\omega_1-\omega_2) $ (Fig.~\ref{fig1}). 
By convention, we define red detuning as $\Delta(R), \delta(R) > 0$.
 The $R$ dependence of the detunings  originates from the $R$-dependent dipole-dipole interaction (DDI) between the molecules, determining their long-range PECs. Additionally, as R varies, the state composition also varies as shown later on in Appendix \ref{tables}, which results in a R dependency on the Rabi frequency. In the following, we thus define $\Omega_1$=$\Omega_1(R \rightarrow \infty)$, $\Omega_2$=$\Omega_2(R \rightarrow \infty)$, $\Delta$=$\Delta(R \rightarrow \infty)$ and $\delta$=$\delta(R \rightarrow \infty)$.
 
\subsection{Adiabatic elimination} \label{Adiabatic elimination}
When $\Delta(R)$ is much larger than $\Omega_1(R)$ and $\Omega_2(R)$, and the radiative decay rate  $\Gamma_q$ of $|q \rangle$, $\tilde{H}^I(R)$ can be reduced, by adiabatic elimination of the $|\tilde{q} \rangle$ state \cite{brion2007}, to the matrix $H^I_{\mathrm{eff}} (R)$ of an effective two-level system expressed in the ($|\tilde{g}_1\rangle, |\tilde{g}_2\rangle$) basis, 
\begin{equation} 
H^I_{\mathrm{eff}} (R) = \hbar
\begin{pmatrix}
0 & -\Omega_{\mathrm{eff}}(R)/2 \\
-\Omega_{\mathrm{eff}}(R)/2 &  \Delta_{\mathrm{eff}}(R)
\end{pmatrix}_{|\tilde{g}_1\rangle, |\tilde{g}_2\rangle},
\label{eq:Heff}
\end{equation}
where 
\begin{equation}
\Omega_{\mathrm{eff}}(R)=\frac{\Omega_1(R) \Omega_2(R)}{2 \Delta(R)},
\label{eq:Omegaeff}
\end{equation}
and
\begin{equation}
\Delta_{\mathrm{eff}}(R)=\delta(R) + \frac{\Omega^2_{1}(R) - \Omega^2_{2}(R)}{4 \Delta(R)}.
\label{eq:deltaeff}
\end{equation}
This scheme is equivalent to the mw shielding (Fig.~ \ref{fig2}), with an important difference: the initial and final states have the same total parity, so that the channels relevant for 2-OS will be different from those of the mw shielding.
Nonetheless, the same requirement still holds: identifying a channel with a repulsive PEC which could be coupled to the entrance channel via a two-photon transition. 
Moreover, it is worth noting that there is no intrinsic limitation on the magnitude of $\Delta_{\mathrm{eff}}$, which could be taken arbitrarily large, as long as it does not reach the next rotational level.

If we now set $\delta(R)=0$ at a given distance, the Raman resonance is achieved so that one of the eigenvectors of the Hamiltonian in Eq.~\ref{eq:Hdressed} is a dark state, $| \Psi_{dark} \rangle =   \left[ \Omega_2(R) | \tilde{g}_1 \rangle - \Omega_1(R) | \tilde{g}_2 \rangle \right]/\sqrt{\Omega_1^2(R) + \Omega_2^2(R)}$, ensuring that the excited state $|q \rangle$ is not populated at this distance, thus exactly cancelling spontaneous emission and photon scattering. Two options can be considered:
\begin{enumerate}
    \item $\delta=0$ for $R \rightarrow \infty$: the individual molecules are then protected against photon scattering, so the ultracold molecular sample in the trap will not be heated while applying the 2-OS scheme (Fig.\ref{fig2}a);
    \item $\delta(R_C)=0$: the molecular pair is less protected against photon scattering while the 2-OS is indeed active at the crossing point between the PECs of the two dressed channels.
\end{enumerate}

At ultracold temperatures, the molecules spend most of their time at very large distances. As argued in \cite{xie2020}, the shielding dynamics proceeds with a characteristic time shorter than the radiative lifetime of a suitable $|q \rangle$ state. Therefore, option (i) is preferable in most experimental realizations. We thus assume $\delta=0$, while the variation of the detunings $\Delta(R)$, $\delta(R)$, and $\Delta_{\mathrm{eff}}(R)$ for the realistic case described below are detailed in appendix \ref{detunings/R}. To ensure a crossing between $|\tilde{g}_1\rangle$ and $|\tilde{g}_2\rangle$ and that the population distribution of the dark state at $R \rightarrow \infty$ is predominantly in $|g_1\rangle$, which is the case for $\Omega_2 > \Omega_1$, we chose $\Delta > 0$ and $\Delta_{\textrm{eff}} < 0$.

\subsection{Basis set and selection rules}

In the presence of an external field, the molecular states $|g_1 \rangle$, $|g_2 \rangle$ and $|q \rangle$ of the pair of identical molecules are appropriately described in the SF frame, and expanded over the basis set of symmetrized vectors ${|[\xi_i,j_i, p_i, \xi_k, j_k, p_k], j_{ik}, \ell , J, M \rangle}$, where $\xi_i$, $j_i$, $p_i$ (resp. $\xi_k$, $j_k$, $p_k$) are the quantum numbers for the electronic state, the rotational state, and the parity of molecule 1 (resp. molecule 2). The angular momenta $\Vec{j}_i$ and $\Vec{j}_k$ are first coupled to yield $\Vec{j}_{ik}$ with the quantum number $j_{ik}$, which is then coupled to the relative angular momentum $\vec{\ell}$ (thus the partial wave $\ell$) to build up the total angular momentum $\vec{J}$ with quantum number $J$ and projection $M$ on the field polarization axis (z axis of the SF frame). We assume that the molecules occupy the lowest vibrational level $v_{i}=v_{k}=0$ of their electronic state, and this label will be omitted in the rest of the paper.

The potential energy operator includes the dominant DDI, which couples basis vectors satisfying $\xi_i \equiv \xi'_i$, $|\Delta j_i| \equiv |j_{i}-j'_{i}|=1$, $p_{i}p'_{i}=-1$, $\xi_k \equiv \xi'_k$, $|\Delta j_k| \equiv |j_{k}-j'_{k}|=1$, $p_{k}p'_{k}=-1$, and $|\Delta \ell| \equiv| \ell-\ell'|=0 ,2$ \cite{lepers2018}. All higher order multipolar interactions are neglected. The diagonalization of this operator yields the PECs schematized in Fig.\ref{fig1}a associated with the eigenvectors $|g_1 \rangle$, $|g_2 \rangle$ and $|q \rangle$, and detailed in section \ref{LR-PECs}. When the shielding lasers are present, they impose selection rules for the one-photon transitions $|g_1 \rangle \rightarrow |q \rangle$, $|q \rangle \rightarrow |g_2 \rangle$, resulting from those on the basis vectors themselves, and depending on the laser polarization. At large distances, the Coriolis effect is small, leaving $\ell$ unaffected by the lasers. We assume that the collision energy is small enough to proceed in the $s$-wave regime ($\ell=0$). The selection rules are:
\begin{itemize}
    \item Circular polarization ($\sigma$): $\Delta \ell=0$, $\Delta J=0 \pm 1$, $\Delta M=+1$ ($\sigma^+$), or $\Delta M=-1$ ($\sigma^-$), and for the quantum numbers of the individual molecules, $\xi_{\alpha} \neq \xi'_{\alpha}$, $|\Delta j_{\alpha}|=1$, $p_{\alpha}p'_{\alpha}=-1$, applying for $\alpha=i$ or $k$, but not both simultaneously.
    \item Linear polarization ($\pi$): $\Delta \ell=0$, $\Delta J=\pm 1$, $\Delta M=0$, and for the quantum numbers of the individual molecules, $\xi_{\alpha} \neq \xi'_{\alpha}$, $|\Delta j_{\alpha}|=1$, $p_{\alpha}p'_{\alpha}=-1$, applying for $\alpha=i$ or $k$, but not both simultaneously.
\end{itemize}

\section{Application to bosonic N\lowercase{a}K}

We apply this proposal to $^{23}$Na$^{39}$K bosonic molecules in the lowest $v_X=0, j_X=0$ rovibrational level of their electronic ground state $X^1\Sigma^+$ (noted $X$ in short afterwards).The molecular hyperfine structure is small \cite{aldegunde2017} and not considered here. In general, the experiments are performed in the presence of an external magnetic field imposed by the location of a Feshbach resonance used to create the ground-state molecules. According to the supplementary material of Ref. \cite{lassabliere2018}, the magnetic field above which the hyperfine structure could be neglected is generally small enough for most alkali-metal diatomic species, opening the possibility to choose a suitable Feshbach resonance. We formulate the 2-OS by relying on the lowest excited electronic state of $^{23}$Na$^{39}$K (the $b^3\Pi_0$ state, noted $b$ afterwards), as in \cite{xie2020}: the bottom of the $b$ PEC lies below all PECs dissociating to the first excited dissociation limit Na($3s$)+K($4p$), so that the two-photon transition could be implemented with a detuning $\Delta$ from the lowest vibrational level $v_{b}=0$. The $b$ state is weakly coupled to the excited $A^1\Sigma^+$ state via spin-orbit interaction (referred to as the $A/b$ system afterwards), yielding a pair of electronic states with $\Omega=0^+$ symmetry, $\Omega$ being the projection of the total electronic angular momentum on the $^{23}$Na$^{39}$K molecular axis. The transition $X \leftrightarrow (A/b)$ is thus dipole-allowed through the $A$ component of the lowest $0^+$ state of the $A/b$ system, resulting in a transition electric dipole moment (TEDM) of 0.0456~a.u. (or 0.116~Debye) \cite{xie2020} for the $v_X=0 \leftrightarrow$ $v_{b}=0$ transition.

\subsection{Long-range potential energy curves}\label{LR-PECs}

The long-range potential energy curves (LR-PECs) are calculated in the SF frame following the same procedure as in \cite{li2019, xie2020}. They include the dominant DDI at first order of perturbation, while the second-order van der Waals terms varying as $R^{-6}$, which results from the presence of the electronically excited states inducing an additional instantaneous small dipole to the system are neglected. We use the same parameters as those reported in \cite{xie2020} NaK: the permanent electric dipole moments (PEDMs) $d_X^0=1.095$~a.u. and $d_b^0=1.220$~a.u, and the rotational constants $B_X^0=0.0950$~cm$^{-1}$ and $B_b^0=0.0951$~cm$^{-1}$ for the $X^1\Sigma^+(v_X = 0)$ and $b^3\Pi_0(v_b = 0)$ levels, respectively, the transition electric dipole moment between these two levels $d_{X \leftrightarrow b}^0=0.0456$~a.u. (all quantities in the body-fixed (BF) frame, and with 1 a.u.=2.541 580 59 Debye). 

Due to the bosonic character of $^{23}$Na$^{39}$K, only even partial waves $\ell=0,2$ are considered when the two molecules are in the $v_X=0,j_X=0$ state, i.e., in the entrance channel with $J=0,2$. We include in our calculations for two ground state molecules colliding with $j_X$ and $j'_X$ even partial waves up to $\ell=8$ to describe states with $j_{i},j_{k}=0,1,2,3,4$, with total angular momentum up to $J=0,1,2$. This yields 22, 33, and 61 basis vectors ${|\xi_i,j_i, p_i, \xi_k, j_k, p_k, j_{ik}, \ell , J, M \rangle}$ for $J=0,1,2$, respectively. As $\ell$ is conserved in our approach, and because the DDI couples states satisfying $\Delta \ell=0,2$, we consider the same range of variation for the quantum numbers when a ground state molecule in a $|X, v_X=0,j_X \rangle$ state collides with another one in a $|b, v_{0^+}=0,j_b \rangle$ state. The size of the basis set is now 119 and 190, for $J=1$ and $J=2$, respectively.

\begin{figure}
    \centering
    \includegraphics[scale=0.8]{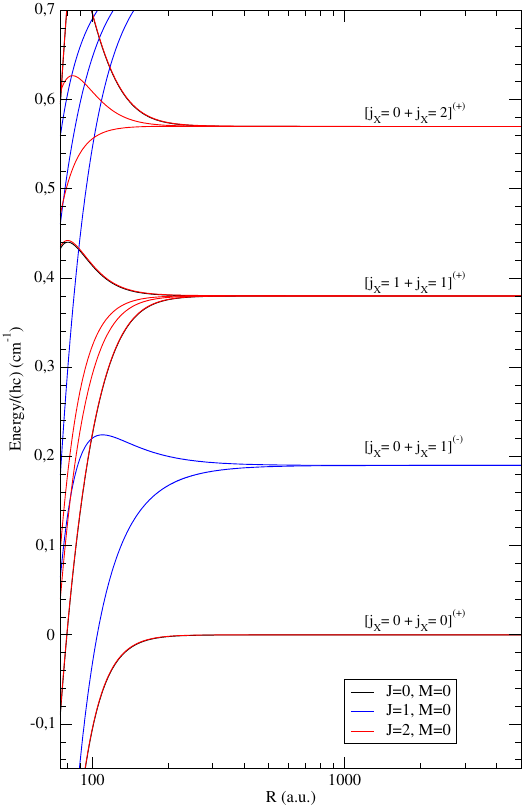} 
    \caption{Adiabatic long-range potential energy curves of two $^{23}$Na$^{39}$K molecules in the $v_X = 0$ level of their electronic X$^{1}\Sigma^{+}$, for the lowest combinations of internal rotational states $j_X$, and for $J=0,2$ with $M=0$.}
    \label{fig:X-Xfig}
\end{figure}

\begin{figure}
    \centering
    \includegraphics[scale=0.8]{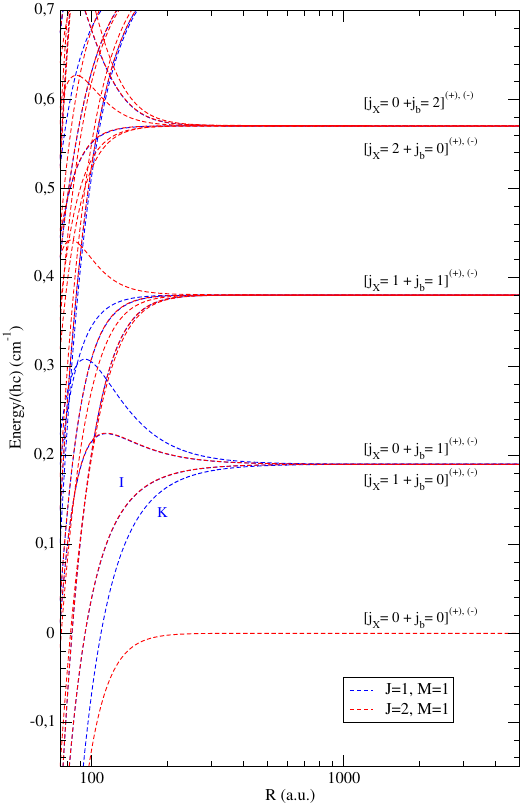} 
    \caption{Adiabatic long-range potential energy curves of two $^{23}$Na$^{39}$K molecules, one in $v_X = 0$, and the other in the level $v_b = 0$ of the $b^3\Pi_0$ state, for the lowest combinations of internal rotational states $j_X$ and $j_b$, and for $J=1,2$ with $M=1$, due to our choice of polarization ($\sigma$) for the OS lasers. Note that the $J=0$ states are not visible as we chose $M=1$. As the molecules are in different states, all curves correspond to degenerate states of + and - parity.}
    \label{fig:X-bfig}
\end{figure}

The LR-PECs for two molecules in $v_X = 0$, or one molecule in $v_X = 0$ interacting with another one in $v_b = 0$, are displayed in Fig.\ref{fig:X-Xfig} and Fig.\ref{fig:X-bfig}, respectively. We note in Fig.\ref{fig:X-bfig} the quasi-degeneracy of the two limits $X(j_X=1) + b(j_b=0)$ and $X(j_X=0) + b(j_b=1)$ resulting from the almost equal rotational constant of the $v_X=0$ and $v_b=0$ levels. This induces a strong mixing between the states correlated to these asymptotes. This pattern is actually present in all heteronuclear alkali-metal diatomic molecules \cite{xie2020}.  Note that the lowest dissociation limit $j_X=0 + j_b=0$ is not appropriate for our 2-OS scheme as it possesses only a $J=0,2$ manifold of states and thus would not be coupled by light to the entrance channel.

\subsection{Mapping the MW-S}

\begin{figure}
\includegraphics[scale=0.47]{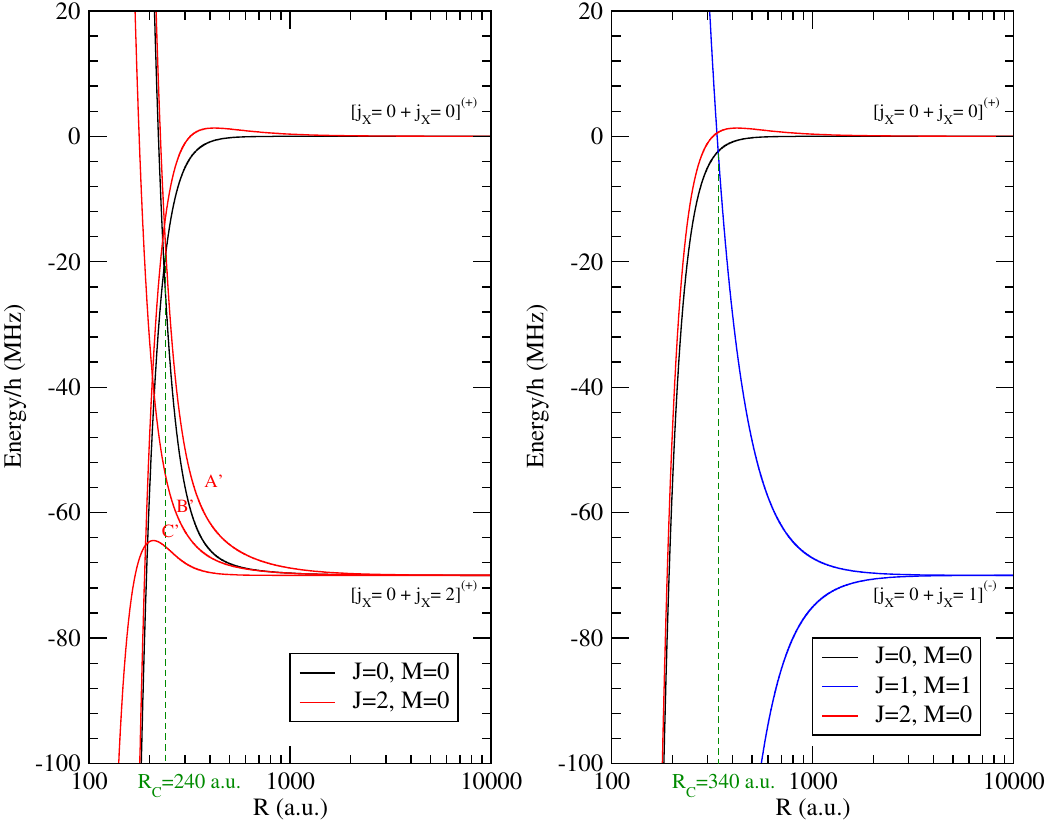}
\caption{(a) The two-optical-photon dressed PECs for two ground state $^{23}$Na$^{39}$K molecules in $R_C\approx240$ a.u., (or $\Delta_{\mathrm{eff}}=-70$~MHz). (b) The one-mw-photon dressed PECs for the same detuning of $\Delta_{\mathrm{mw}}=-70$~MHz following \cite{lassabliere2018}. }
\label{fig:2OS-MW}
\end{figure}

In order to map the mw shielding scheme, we build the 2-OS scheme by choosing the $|g_2\rangle$ state among the molecular pair states correlated to the $[j_X=0+j_X=2]^{(+)}$, which exhibits several repulsive LR-PECs (the A', B' and C' curves in Fig. \ref{fig:2OS-MW}).The components of these adiabatic states associated with the three repulsive LR-PECs A', B', C' (for $J=2$), are reported in Table \ref{tab:limiteX-X-prime} of appendix \ref{tables}, expressed on a basis set limited to 15 vectors.Note that these curves varies as $C_6/R^6$ with "giant" $C_6$ coefficients, with magnitude of about $-10^6$a.u. for the A' curve, and a few $-10^5$a.u. for the B' curve (see for instance Refs. \cite{lepers2018,lepers2013,vexiau2015}). In the entrance channel we calculated the LR-PECs for $J=0,2$. The $J=2$ curve is entirely determined by the centrifugal barrier $\ell=2$ with a height of about $k_B \times 63.26$ $\mu$K, much higher than the typical collision energy in ongoing cold molecule experiments ($E_{col}/k_B$ of a few hundreds of nK). We thus only considered the channel $J=0$ as $|g_1 \rangle$, composed of a single basis vector. To comply with the selection rules, the $|q\rangle$ state is taken from the set of adiabatic states correlated to the $[j_X=0+j_b=1]^{(-)} $ and $[j_X=1+j_b=0]^{(-)} $ quasi-degenerate dissociation limits. The adiabatic states, and the related PECs, are obtained from the diagonalization of the DDI in the basis sets above for every intermolecular (large) distance $R$. The coupling of the adiabatic states by the OS lasers is determined by the non-zero matrix elements of the $X\leftrightarrow b$ TEDM between basis vectors fulfilling the selection rules.  After eliminating the intermediate $|q \rangle$ state, the 2-OS scheme results in a crossing point $R_C$ between an attractive entrance PEC and a repulsive rotationally excited one, corresponding to a given value of the effective detuning $\Delta_\mathrm{eff}$.

We illustrate the selected LR-PECs for the 2-OS of collisions between two ground state $^{23}$Na$^{39}$K molecules in Fig.\ref{fig:2OS-MW}a. Assuming an adiabatic connection of $|g_1 \rangle$, $|g_2 \rangle$ and $|q \rangle$ to their dissociation limit, the 2-photon transition can be labeled as $|g_1 \rangle[j_X=0+j_X=0]^{(+)} \rightarrow |q \rangle[j_X=0+j_b=1]^{(-)} \rightarrow |g_2 \rangle[j_X=0+j_X=2]^{(+)}$. After investigating the composition of all the adiabatic states on the above basis set and by taking into account the selection rules, we can characterize an efficient scheme with a pair of $\sigma$ polarized photons connecting the appropriate components (listed in Table \ref{tab:2OS-components70MHz} for $\Delta_{\mathrm{eff}}=-70$ MHz) of $|g_1 \rangle$ to those of the I and K excited states, and in turn to those of the A', B', and C' repulsive states (Fig.\ref{fig:2OS-MW}a).  In appendix \ref{tables}, we display a similar table for other values of $\Delta_{\mathrm{eff}}$ (or $R_C$ values), showing that the variation in $R$ of the composition of the adiabatic states is weak enough to keep the same character at all distances larger than $R_C$. Equivalently, a pair of $\pi$ polarized photons can be used as it also fulfills the selection rules, with the appropriate change of angular factors.

%%%%%%%%%%%%%%%%%%%%%%%%%%%%%%%%%%%%%%%%%%%%%%%%%%%%%%%%%%
%\begin{table} 
%\centering
%\begin{tabular}{|c|c|c|}
%\hline
%State & ${|[\xi_i,j_i,p_i,\xi_k,j_k,p_k],j_{ik},\ell,J,M\rangle}$& Component  \\
%\hline 
%\hline
%$|g_1\rangle$   &$|[X,0,1,X,0,1]  ,0,0,0,0 \rangle$& 99.95\% \\ 
%\hline
%$|q \rangle$(I) &$|[X,0,1,b,1,-1] ,1,0,1,1 \rangle$&  32.74\%  \\
%$|q \rangle$(K) &$|[X,0,1,b,1,-1] ,1,0,1,1 \rangle$&  16.69\%  \\
%\hline
%\hline
%$|g_2\rangle$(A')&$|[X,0,1,X,2,1],2,0,2,0 \rangle$&5.48\% \\
% \hline
%$|g_2\rangle$(B')&$|[X,0,1,X,2,1],2,0,2,0 \rangle$&0.48\% \\
% \hline
%  $|g_2\rangle$(C')&$|[X,0,1,X,2,1],2,0,2,0 \rangle$&93.95\% \\
% \hline
%\end{tabular}
%\caption{The largest components (in \%) of the $|g_1 \rangle$, $|g_2 \rangle$ and $|q \rangle$ adiabatic states fulfilling the selection rules for the two-photon transition in $\sigma^+ -  \sigma^+$ polarization at $R_C \approx 293.75$ a.u., assuming $\Delta_{\mathrm{eff}}=-70$~MHz.}
%\label{tab:2OS-components70MHz}
%\end{table} 

\begin{table}
\centering
\begin{tabular}{|c|c|c|}
\hline
State & ${|[\xi_i,j_i,p_i,\xi_k,j_k,p_k],j_{ik},\ell,J,M\rangle}$& Component  \\
\hline 
\hline
$|g_1\rangle$   &$|[X,0,1,X,0,1]  ,0,0,0,0 \rangle$& 99.95\% \\ 
\hline
$|q \rangle$(I) &$|[X,0,1,b,1,-1] ,1,0,1,1 \rangle$&  33.06\%  \\
$|q \rangle$(K) &$|[X,0,1,b,1,-1] ,1,0,1,1 \rangle$&  16.74\%  \\
\hline
$|g_2\rangle$(A')&$|[X,0,1,X,2,1],2,0,2,0 \rangle$&10.90\% \\
 \hline
$|g_2\rangle$(B')&$|[X,0,1,X,2,1],2,0,2,0 \rangle$&9.89\% \\
 \hline
  $|g_2\rangle$(C')&$|[X,0,1,X,2,1],2,0,2,0 \rangle$&78.88\% \\
 \hline
\end{tabular}
\caption{The components (in \%) of the $|g_1 \rangle$, $|g_2 \rangle$ and $|q \rangle$ adiabatic states which fulfill the selection rules for the two-photon transition in $\sigma$ polarization at $R_C \approx 240$~a.u., assuming $\Delta_{\mathrm{eff}}=-70$~MHz.}
\label{tab:2OS-components70MHz}
\end{table} 

By comparing panels (a) and (b) in Fig. \ref{fig:2OS-MW}, we realize that our 2-OS scheme nicely maps onto the one-mw-photon shielding approach proposed in \cite{lassabliere2018,karman2018} based on the mw transition $j_X=0 \rightarrow j_X=1$, and observed experimentally with $^{23}$Na$^{40}$K molecules with $\Delta_{\mathrm{mw}}=-2 \pi \times 8$~MHz and  $\Omega_{\mathrm{mw}}=2 \pi \times 11$~MHz \cite{schindewolf2022}. Choosing here the same values for $\Delta_{\mathrm{eff}}$ and $\Omega_{\mathrm{eff}}$, $\Omega_1$ and $\Omega_2$ can be evaluated for different values of $\Delta$ and $\delta$ as $\Omega_2= 2 \Delta \Omega_{\mathrm{eff}}/ \Omega_1$ and 
\begin{equation}
\label{omega_1}
\Omega_1= \sqrt{2 \Delta} \sqrt{(\Delta_{\mathrm{eff}} - \delta)+ \sqrt{(\Delta_{\mathrm{eff}}-\delta)^2+\Omega_{\mathrm{eff}}^2}}.
\end{equation}

\section{Discussion and conclusion}

\begin{figure}[h!]
    \includegraphics[scale=0.55]{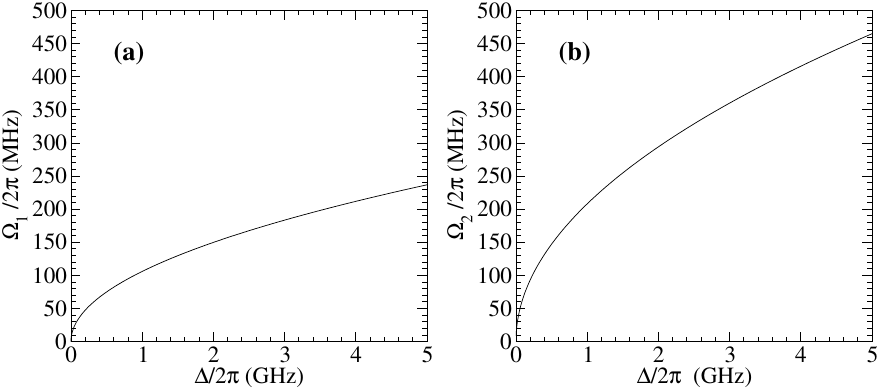}
    \caption{The one-photon Rabi frequency (a) $\Omega_1/2 \pi$ and (b) $\Omega_2/2 \pi$ as functions of the one-photon detuning $\Delta/2 \pi$, with $\delta=0$, $\Omega_{\mathrm{eff}}/2 \pi=  11$~MHz and $\Delta_{\mathrm{eff}}/2 \pi= - 8$~MHz, identical to the experimental values of \cite{schindewolf2022} for the mw shielding. }
    \label{fig:omega1omega2}
\end{figure}

The working principle of 2-OS and its mapping onto MW-S is best explained by the “cleanest” coupling of the (weakly dipole-allowed) $X$-$b$ transition: the $b$ state is the lowest electronic state accessible via a dipole-allowed transition, leading to the most isolated, or closest to the idealized 3-level scheme. Considering the weak TEDM of the quasi-forbidden one-photon transition $v_X=0 \rightarrow v_{b(0^+)}=0$, the range of $\Omega_1$ and $\Omega_2$ values varying with $\Delta$ (with $\delta=0$) displayed in Fig.\ref{fig:omega1omega2}), leads to quite large laser intensities. However, as the Raman condition $\delta=0$ is fulfilled at infinity (\textit{i.e.} for individual molecules), one can significantly reduce the value of $\Delta$ without hampering the efficiency of the 2-OS scheme. Moreover, as the proposed scheme can be generalized to any intermediate state $|\tilde{q} \rangle$, excited electronic states with large TEDM can be readily used, allowing for moderate laser intensities $I_1$ and $I_2$. For example, coupling via the $A^1 \Sigma^+$ state in $^{23}$Na$^{39}$K yields the desired $\Omega_{\mathrm{eff}}/(2 \pi) \sim 10$~MHz with an estimated $I_{2}~(\gg I_{1})$ on the order of $10^{4}$ W/cm$^2$ for $\Omega_1 / \Omega_2 \sim 10^{-2}$, and a loss probability due to off-resonant scattering below $10^{-6}$ per collision. Note that the two-photon coupling scheme allows using a common laser source in combination with frequency modulation for driving both optical transitions. Thus, the optical phase noise is common mode and the overall phase stability is determined by the purity of the microwave modulation source. Therefore, we expect that the proposed scheme could be successfully implemented in a forthcoming experiment, helped by full dynamical calculations extending those of Ref. \cite{xie2020} to the 2-OS case.

In summary, we have proposed a new scheme for shielding of inelastic and reactive short-range collisions based on two-photon transitions. It allows taking advantage of optically driven transitions including insensitivity to polarization and flexibility in the choice of electronic states, while suppressing undesired off-resonant photon scattering which was present in the previously proposed 1-OS. Our method is applicable to a broad range of bialkali molecules, with expected efficiencies comparable to the previously demonstrated mw shielding scheme. Our results may be of importance in experiments where collisional losses in general pose a major limitation to the achievable lifetimes and densities of ultracold molecular gases.

\section{Acknowledgments}
C.K. acknowledges the support of the Quantum Institute of Université Paris-Saclay. M.M., S.O., and L.K. thank the DFG (German Research Foundation) for support through CRC 1227 DQ-mat and Germany’s Excellence Strategy— EXC-2123 QuantumFrontiers—No. 390837967. This work is supported in part by the ERC Consolidator Grant 101045075- TRITRAMO, and by the joint ANR/DFG project OpEn375 MInt (ANR-22-CE92-0069-01). Stimulating discussions with Prof. Eberhard Tiemann (IQO, Leibniz University, Hannover) and with Dr Patrick Cheinet (LAC, CNRS, Université Paris378 Saclay, France) are gratefully acknowledged.

%\appendix

\section{Appendices}

%\chapter{Variation of the Rabi frequencies for the 2-OS}

\subsection{Variation of the Rabi frequencies for the 2-OS}
\begin{figure}[h]
    \centering
    \includegraphics[scale=0.6]{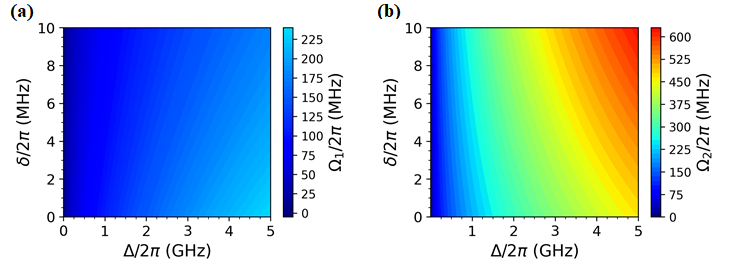}
    \caption{The one-photon Rabi frequency (a) $\Omega_1/2 \pi$ and (b) $\Omega_2/2 \pi$ as functions of the one-photon detuning $\Delta/2 \pi$, and of $\delta/2 \pi$, with $\Omega_\mathrm{eff}/2 \pi= 11$~MHz and $\Delta_\mathrm{eff}/2 \pi= - 8$~MHz, identical to the experimental values of \cite{schindewolf2022} for the mw shielding.}
    \label{fig:omega1omega2-color}
\end{figure}

In Fig.\ref{fig:omega1omega2-color} we present a generalization of Fig.\ref{fig:omega1omega2}, showing the variation of the individual Rabi frequencies (a) $\Omega_1$ and (b) $\Omega_2$ as functions of $\Delta$ and $\delta$, for fixed values of $\Omega_\mathrm{eff}$ and  $\Delta_\mathrm{eff}$. We see that the variation with $\delta$ is smooth, so that the experimental realization of the 2-OS could consider some flexibility on $\delta$, with a possible compromise between the 2-OS efficiency versus the full cancellation of the photon scattering rate.

\subsection{Variation of the detunings characterizing the 2-OS}\label{detunings/R}

\begin{figure}[b!]
    \centering
    \includegraphics[scale=0.825]{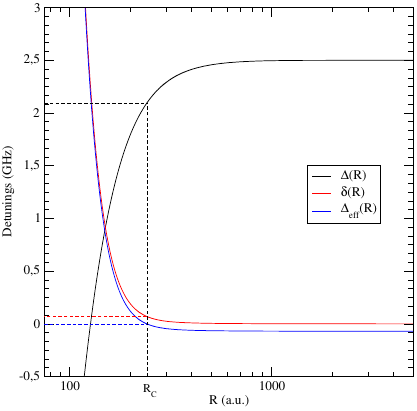}
    \caption{Variation of $\Delta(R)$, $\Delta_\mathrm{eff}(R)$ and $\delta(R)$ with the intermolecular distance $R$, fixing $\Delta=2.5$~GHz and $\delta=0$ for $R \rightarrow \infty$ (case (i) in the main text), and $\Delta_\mathrm{eff}=-70$~MHz.  The dashed blue line locates the crossing point $R_C \approx 240$~a.u. between the dressed scattering channels, where $\Delta_\mathrm{eff}(R_C)=0$. The dashed black line shows the value of $\Delta(R_c)=2.09$~GHz.}
    \label{fig:detuning_no_raman}
\end{figure}

In Section \ref{Interaction Hamiltonian}, we exposed the reasons for the dependence of the detunings $\Delta$, $\delta$, and $\Delta_\mathrm{eff}$ on the intermolecular distance $R$, \textit{i.e.}, as the collision between the molecules develops. These variations are displayed in Fig. \ref{fig:detuning_no_raman} when the Raman condition $\delta=0$ is fulfilled at infinity, thus for individual molecules (case (i) in Section \ref{Adiabatic elimination}). We see that the $R$ variation of the PEC of the $|q\rangle$ states, and thus of $\Delta(R)$, does not affect the behaviour of the 2-OS: namely, the variations of $\Delta_\mathrm{eff}$ and $\delta$ remain weak for $R>R_C$, and the magnitude of $\Delta$ dominates the process for all distances. In other words, the conditions for adiabatic elimination remain fulfilled at all relevant distances.

\subsection{Variation of the adiabatic states composition for the 2-OS}\label{tables} 
In Table \ref{tab:main-components}, we report, for several crossing distances $R_C$ - or effective detunings $\Delta_\mathrm{eff}$-, the largest components of the I and K excited states that can be coupled to the entrance channel, and the largest components of the A', B', and C' states that can be coupled to those I and K components. We see that the variation of the composition of the adiabatic states is smooth over a broad range of distances, yielding flexibility to the choice of laser frequencies and intensities which, for a constant target shielding efficiency($\Omega_\mathrm{eff}/2 \pi= 11$~MHz and $\Delta_\mathrm{eff}/2 \pi= - 8$~MHz), allows balancing between an optimal suppression of residual off-resonant scattering and a favorable composition of the dark state (that is, minimizing the $[X(j_X=0)+X(j_X=2)^{(+)}]$ component).

\begin{table*}[tp]
\centering
\begin{tabular}{|c|c|c|c|c|}
\hline
Adiabatic state & Basis vector &  &   &    \\
&${|[\xi_i,j_i, p_i, \xi_k, j_k, p_k], j_{ik}, \ell , J, M \rangle}$&
 $\Delta_\mathrm{eff}= -100$MHz &$\Delta_\mathrm{eff}= -500$MHz& $\Delta_\mathrm{eff}= -1$GHz  \\
\hline
\hline
 &  & $R_C= 225.2$ a.u. & $R_C=167.9$ a.u.& $R_C=148$ a.u. \\
\hline
\hline
Entrance state  & $|[X,0,1,X,0,1],0,0,0,0 \rangle$  & 99,75\%       & 98.61\%     & 97.11\% \\ 
\hline
\hline
A' & $|[X,0,1,X,2,1],2,0,2,0 \rangle$ & $12.40\% $ & $16.16\%$ &$ 16.14\% $  \\
\hline
B' & $|[X,0,1,X,2,1],2,0,2,0 \rangle$ & $15.30\% $& $ 32.11\%$ & $34.17\% $  \\
\hline
C' & $|[X,0,1,X,2,1],2,0,2,0 \rangle$ & $71.82\% $& $49.26\%$ & $44.91\% $  \\
\hline
\hline

I & $|[X,0,1,b,1,-1] ,1,0,1,1 \rangle $ & 33.12\% & 33.17\% & 33.04\%  \\
\hline
K & $|[X,0,1,b,1,-1] ,1,0,1,1 \rangle $ & 16.74\% & 16.68\% &  16.58\%  \\

\hline

\end{tabular}
\caption{The largest components (in \%) on the ${|[\xi_i,j_i, p_i, \xi_k, j_k, p_k], j_{ik}, \ell , J, M \rangle}$ basis set, of the I and K excited states (Fig.\ref{fig:X-bfig}) that can be coupled to the $|[X,0,1,X,0,1],0,0,0,0 \rangle$ entrance channel (first line), and of the A', B', and C' states(Fig. \ref{fig:2OS-MW}) that can be coupled to the above I and K components.}
\label{tab:main-components}
\end{table*}

We also present in tables \ref{tab:limiteX-X-prime} and \ref{tab:limiteX-b} the full composition of the adiabatic states at a Condon point $R_C=167.9$a.u. corresponding to an effective detuning of $\Delta_\mathrm{eff}=-500$ MHz.

\begin{table}[h]
\centering
\begin{tabular}{|c|c|c|c|}
\hline
Basis vector &  \multicolumn{3}{|c|}{Component} \\
${|[\xi_i,j_i,p_i,\xi_k,j_k,p_k],j_{ik},\ell\rangle}$ &\multicolumn{3}{|c|}{} \\
\hline \hline
J=2, M=0 &  A' & B' & C'\\
\hline
$|[X,0,1,X,2,1],2,0 \rangle $ &\textbf{ 16.16 \% }& \textbf{32.11 \%}& \textbf{49.26 \%}\\
$|[X,0,1,X,2,1],2,2 \rangle $ & 24.71 \% & 60.75 \% &11.36 \%\\
$|[X,0,1,X,2,1],2,4 \rangle $ & 53.11 \% &4.82 \% &38.56\% \\
$|[X,1,-1,X,1,-1],0,2 \rangle $ & 3.8 \% &$<$ 0.01 \% & 0.01 \%\\
$|[X,1,-1,X,1,-1],2,0 \rangle $ & 0.38 \% &0.93 \% & 0.18 \%\\
$|[X,1,-1,X,1,-1],2,2 \rangle $ & 0.54 \% &1.15 \% & 0.17 \%\\
$|X,1,-1,X,1,-1],2,4  \rangle $ & 1.06 \% &0.04 \% & 0.22 \%\\
$|[X,1,-1,X,3,-1],2,0 \rangle $ & $<$ 0.01 \% & $<$ 0.01 \% & $<$ 0.01 \%\\
$|[X,1,-1,X,3,-1],2,2 \rangle $ & $<$ 0.01 \% & $<$ 0.01 \% & $<$ 0.01 \%\\
$|[X,1,-1,X,3,-1],2,4 \rangle $ & $<$ 0.01 \% & $<$ 0.01 \% & $<$ 0.01 \%\\
$|[X,1,-1,X,3,-1],3,2 \rangle $ & $<$ 0.01 \% & 0.05 \% & 0.01 \%\\
$|[X,1,-1,X,3,-1],3,4 \rangle $ & $<$ 0.01 \% & 0.02 \% & 0.01 \%\\
$|[X,1,-1,X,3,-1],4,2 \rangle $ & 0.05 \% & 0.01 \% & 0.1\%\\
$|[X,1,-1,X,3,-1],4,4 \rangle $ & 0.05 \% & 0.07 \% & $<$ 0.01 \%\\
$|[X,1,-1,X,3,-1],4,6 \rangle $ & 0.1 \% & 0.01 \% & 0.08\%\\
\hline
\hline
\end{tabular}
- \caption{The components (in \%) of the A', B' and C' adiabatic states, evaluated for $\Delta_\mathrm{eff} = - 500$MHz ($R_C=167.9$a.u.). The numbers in boldface correspond to the component on the vector which is allowed by the selection rules.}
\label{tab:limiteX-X-prime}
\end{table}

\begin{table}[h]
\centering
\begin{tabular}{|c|c|c|}
\hline
Basis vector & \multicolumn{2}{|c|}{Component} \\
${|[\xi_i,j_i,p_i,\xi_k,j_k,p_k],j_{ik},\ell\rangle}$ & \multicolumn{2}{|c|}{} \\
\hline \hline
$J=1$, $M=1$ &  I &  K\\
\hline
$|[X,0,1,b,1,-1],1,0 \rangle $ & \textbf{33.16 \%} & \textbf{16.68 \%}  \\
$|[X,0,1,b,1,-1],1,2 \rangle $ & 16.35 \% & 32.92\%  \\
$|[X,1,-1,b,0,1],1,0 \rangle $ & 33.49 \% & 16.76 \%  \\
$|[X,1,-1,b,0,1],1,2 \rangle $ & 16.55 \% & 33.08\%  \\
$|[X,1,-1,b,2,1],1,0 \rangle $ & $<$ 0.01 \% & $<$ 0.01 \% \\
$|[X,1,-1,b,2,1],1,2 \rangle $ & $<$ 0.01 \% & $<$ 0.01 \%\\
$|[X,1,-1,b,2,1],2,2 \rangle $ & 0.06\% & $<$ 0.01 \% \\
$|[X,1,-1,b,2,1],3,2 \rangle $ & 0.09\% & 0.11\% \\
$|[X,1,-1,b,2,1],3,4 \rangle $ & 0.06\% & 0.15 \%\\
$|[X,2,1,b,1,-1], 1,0 \rangle $ & $<$ 0.01 \% &$<$ 0.01 \% \\
$|[X,2,1, b,1,-1],1,2 \rangle $ & $<$ 0.01 \% & $<$ 0.01 \%\\
$|[X,2,1,1,-1],2,2 \rangle $ & 0.06 \% & $<$0.01\% \\
$|[X,2,1,b,1,-1],3,2 \rangle $ & 0.1\% & 0.11\% \\
$|[X,2,1,b,1,-1],3,4 \rangle $ & 0.07 \% & 0.15\%\\
\hline

\end{tabular}
\caption{ The components (in \%) of the I and K adiabatic states, evaluated for $\Delta_\mathrm{eff} = -500$MHz or $R_C=167.9$a.u.. The numbers in boldface correspond to the components on the vectors which are allowed by the selection rules. Note that the states connected to $[j_X=0+j_b=1]^{(-)}$ are strongly mixed to $[j_X=1+j_b=0]^{(-)}$ by the DDI.}
\label{tab:limiteX-b}
\end{table}

\bibliographystyle{ieeetr}

\end{document}